\documentclass[preprintnumbers,showpacs,showkeys,amsmath,amssymb,floatfix,prd,
superscriptaddress,nofootinbib]{revtex4}

\usepackage{latexsym}
\usepackage{epsfig}
\usepackage{amssymb}
\usepackage{graphicx}

\begin{document}

\title{An electromagnetic extension of the Schwarzschild interior solution and the corresponding Buchdahl limit}

\author{Ranjan Sharma}
\email{rsharma@associates.iucaa.in} \affiliation{Department of Physics, Cooch Behar Panchanan Barma University, Cooch Behar 736101, India.}

\author{Naresh Dadhich}
\email{nkd@iucaa.in}
\affiliation{The Inter-University Centre for Astronomy and Astrophysics, Post Bag 4, Pune 411007, India.}

\author{Shyam Das}
\email{dasshyam321@gmail.com}
\affiliation{Department of Physics, P. D. Women's College, Jalpaiguri 735101, India.}

\author{Sunil D. Maharaj}
\email{maharaj@ukzn.ac.za}
\affiliation{Astrophysics and Cosmology Research Unit, School of Mathematics, Statistics and Computer Science, University of KwaZulu-Natal,
Private Bag X54001, Durban 4000, South Africa.}

\begin{abstract}
We wish to construct a model for charged star as a generalization of the uniform density Schwarzschild interior solution. We employ the Vaidya and Tikekar ansatz [{\it Astrophys. Astron.} {\bf 3} (1982) 325] for one of the metric potentials and electric field is chosen in such a way that when it is switched off the metric reduces to the Schwarzschild. This relates charge distribution to the Vaidya-Tikekar parameter, $k$, indicating deviation form sphericity of three dimensional space when embedded into four dimensional Euclidean space. The model is examined against all the physical conditions required for a relativistic charged fluid sphere as an interior to a charged star. We also obtain and discuss charged analogue of the Buchdahl compactness bound.
\end{abstract}

\maketitle


\section{Introduction}
\label{sec1}
In 1916, almost immediately after the derivation of the unique exact solution describing the exterior gravitational field of a static spherically symmetric isolated object, Schwarzschild \cite{Sch1,Sch2} obtained an interior solution for a uniform density fluid sphere of finite radius whose exterior was described by the former metric. Since then, a large number of exact solutions have been obtained for a more realistic description of relativistic compact stars, out of which only a few are shown to be physically viable, regular and well-behaved \cite{Delgaty}. A natural extension of such models has been the inclusion of the electromagnetic field. The corresponding Einstein-Maxwell system, being highly non-linear, is extremely difficult to solve, and hence different simplifying techniques are often invoked to solve such a system. The choice of the Vaidya and Tikekar (VT) \cite{Vaidya} metric ansatz, is one such approach which, ever since its inception,  has found huge success for modelling of realistic astrophysical objects. The VT model was subsequently generalized by many investigators \cite{Knutsen87,Tikekar90,leach1,sailo1,Thiru19}. This ansatz is motivated by a geometric property that $t= const.$ hypersurface of the associated spacetime, when embedded in a $4$-Euclidean space is not spherical but spheroidal. The parameter $k$, which appears in the ansatz, indicates the departure from the sphericity of associated $3$-space. The VT ansatz has been utilized to model a wide variety of compact stars (see e.g., Ref.~\cite{herx1,sax,Sharma07,kar1,koma1,jit1,Chatto,Paul2014}) and radiating stars (see e.g., Ref.~\cite{RSRT,Vaidya4,Sarwe,Sharma1}). The model has found its applications in higher dimension spacetimes as well (see, e.g., Ref.~\cite{Khug}). Recently an anisotropic stellar model has also been developed in the Buchdahl-Vaidya-Tikekar metric ansatz \cite{Maurya19}.

In this paper, we revisit the Vaidya-Tikekar metric ansatz to model a static spherically symmetric compact star with a charged fluid interior. Incorporation of the electromagnetic field in the modelling of relativistic astrophysical objects has a long history  and is relevant for a wide variety of astrophysical systems at different stages of its evolution. Some of the pioneering works in this field include the investigations of Majumdar \cite{Majumdar}, De and Raychaudhari \cite{De}, Papapetrou \cite{Papa}, Cooperstock and Cruz \cite{Cooper} and Bekenstein \cite{Beken}, to name a few. Among many other factors, the observed high value of the electric field at the surface of ultra-compact stars \cite{Usov} provides a sufficient ground to study the Einstein-Maxwell system. A large class of analytic solutions to the Einstein-Maxwell system corresponding to the exterior Reissner-Nordstr\"om metric has been compiled by Ivanov \cite{Ivanov}.

There has been considerable work in the literature on charged fluid model with the VT ansatz and with different prescriptions for the electric field, and models have been examined for their physical viability and acceptability \cite{Sharma2001,Tikekar84,Patel87,koma1,Kum18}. To get a solution of the Einstein-Maxwell field equations, one has always to prescribe some fall off behaviour for electric field and/or choose a specific value of the spheroidal parameter $k$. All these papers and the references given therein offer a good spectrum of different choices. In this paper, our main motivation is to superimpose electric field distribution on uniform density fluid distribution. We demand that when the electric field is switched off, the solution should reduce to the uniform density Schwarzschild solution. This limit would require the parameter $k \to 0$, which means electric charge should be proportional to geometric parameter $k$. This, of course, amounts to prescribing a particular charge and electric field distribution. The novel and interesting feature of this model is clubbing of $k$ with charge distribution. It is matched to R-N metric at the boundary defined by $p=0$, and the solution is physically fully viable and satisfactory for modelling an astrophysical charged stellar object.

We obtain and discuss charged analogue of the Buchdahl limit \cite{Buch} for the model. In relativistic astrophysics, measurement of the maximum mass to radius ratio of a stellar configuration has been a matter of prime importance ever since Buchdahl \cite{Buch} derived the relation  $M/R \leq 4/9$ for an isotropic fluid sphere. In a very simple manner the Buchdahl bound could be found by demanding pressure at the centre being finite. This is what has motivated us to seek charged analogue of the uniform density Schwarzschild fluid sphere so that we could obtain the charged analogue of the Buchdahl limit. Theoretical developments in the analysis of Buchdahl limit for Schwarzschild as well as Reissner-Nordstr\"om background spacetimes is available in Ref.~\cite{Dadhich20} and references therein. In this paper, by restricting the model parameters in such a manner that central pressure does not diverge, we obtain an upper bound on compactness of charged star which is a charged generalization of the Buchdahl limit.

The paper is organized as follows. In section \ref{sec2}, for a static and spherically charged fluid distribution, we lay down the independent set of equations for the Einstein-Maxwell system in terms of a single generating function $f(r)$. The idea of this generating function is such that by setting $f(r) =0$, one gets back the Schwarzschild solution for a homogeneous fluid sphere. For $f(r) \neq 0$, we specify the charge distribution $q(r)$ so as to have fallen back on the Schwarzschild solution. The most interesting choice we would explore is $f(r) = k r^2/C^2$, which is, in fact, the VT-metric ansatz, and we generate a new class of solutions where the electromagnetic field gets determined in terms of the parameter $k$. In section  \ref{sec3}, we match the interior solution to the exterior Reissner-Nordstr\"om metric at the boundary given by $p=0$. Making use of the junction conditions, we determine the constants of the model in terms of total mass $M$, radius $R$ and charge $Q$. In section \ref{sec4}, we discuss the Buchdahl limit, some physical properties and overall viability of the model. We end with a discussion in section \ref{sec5}.

\section{Einstein-Maxwell system}
\label{sec2}
We write the line-element describing the interior of a static spherically symmetric charged fluid sphere in the form
\begin{equation}
ds_{-}^2 = -e^{2\nu} dt^2 + e^{2\mu}dr^2 + r^2(d\theta^2 + \sin^2\theta d\phi^2),\label{intm1}
\end{equation}
in standard coordinates $x^i = (t, r, \theta, \phi)$ where $\nu(r)$ and $\mu(r)$ are the undetermined functions. The unknown functions can be determined by solving the Einstein-Maxwell field equations
\begin{eqnarray}
G_{ij} &=& R_{ij}-\frac{1}{2}g_{ij}R = {8\pi}(T_{ij}+E_{ij}), \label{efe1}\\
F_{[ij,k]} &=& 0, ~~~~~\left(e^{-(\nu+\mu)}r^2E\right)' =  {4\pi}\sigma e^{\mu} r^2,\label{max1}
\end{eqnarray}
where,
\begin{eqnarray}
T_{ij} &=& (p+\rho)u_iu_j + pg_{ij},\label{em1}\\
E_{ij} &=& \frac{1}{4\pi}(F_i^l F_{jl}-\frac{1}{4}g_{ij} F^{lm}F_{lm}),\label{em2}
\end{eqnarray}
represent the energy-momentum tensor corresponding to matter and electromagnetic field, respectively. In Eqs.~(\ref{em1}) and (\ref{em2}), $\rho$, $p$, $\sigma$ and $F_{ij}$ denote the energy-density, pressure, charge-density and electromagnetic field tensor, respectively. $u^{i}$ is the $4$-velocity of the fluid. Due to spherical symmetry, the only surviving independent component of the electromagnetic field tensor is $F_{tr} E$. The Maxwell equations (\ref{max1}) yield
\begin{equation}
E = \frac{e^{(\nu+\mu)}}{r^2}q(r),\label{em3}
\end{equation}
where the total charge $q(r)$ contained within the sphere of radius $r$ is defined as
\begin{equation}
q(r) = 4\pi \int_0^r\sigma r^2 e^\mu dr.\label{chareq}
\end{equation}
From Eq.~(\ref{em3}), we have
\begin{equation}
F^2 = F_{tr}F^{tr} = -e^{-2(\nu+\mu)} E^2 =  -\frac{q^2(r)}{r^4}.
\end{equation}
The Einstein-Maxwell field equations (in system of units having $G = c = 1$) are then obtained as
\begin{eqnarray}
{8\pi}\left(\rho +\frac{q^2}{8\pi r^4}\right) &=& \frac{(1-e^{-2\mu})}{r^2}+\frac{2\mu' e^{-2\mu}}{r},\label{e1}\\
{8\pi}\left(p-\frac{q^2}{8\pi r^4}\right) &=& \frac{2\nu' e^{-2\mu}}{r}-\frac{(1-e^{-2\mu})}{r^2},\label{e2}\\
{8\pi}\left(p+\frac{q^2}{8\pi r^4}\right) &=& e^{-2\mu}\left(\nu''+\nu'^2-\nu'\mu'+\frac{\nu'}{r}-\frac{\mu'}{r}\right)\label{e3},\\
4\pi\sigma&=&\frac{e^{-\mu}}{r^2}\frac{dq}{dr},\label{e4}
\end{eqnarray}
where, a prime ($'$) denotes differentiation with respect to the radial parameter $r$. Combining equations (\ref{e2}) and (\ref{e3}), we obtain
\begin{equation}
\frac{2q^2}{r^4} = e^{-2\mu} \left(\nu''+\nu'^2-\nu'\mu'-\frac{\nu'}{r}-\frac{\mu'}{r}-\frac{1-e^{2\mu}}{r^2}\right),\label{pie}
\end{equation}
which becomes the definition of the electric field provided $\mu(r)$ and $\nu(r)$ are known. To solve the system, we assume the metric potential $\mu(r)$ in the Buchdahl-VT \cite{Vaidya,Buch} form as
\begin{equation}
e^{\mu} = \sqrt{\frac{1+ f(r)}{1 - \frac{r^2}{C^2}}}\label{lm2},
\end{equation}
where $C$ is an arbitrary constant. It is not yet the Schwarzschild solution because $\nu$ is still undetermined. Note that when $f(r)=0$, one can generate the Schwarzschid interior solution for an incompressible fluid sphere.

A coordinate transformation $x^2=1-\frac{r^2}{C^2}$ allows us to rewrite Eq.~(\ref{pie}) in the form
\begin{equation}
\frac{d^2\nu}{d x^2}+\left(\frac{d\nu}{dx}\right)^2-\left[\frac{f_x}{2(1+f)}\right]\frac{d\nu}{d x} +\frac{xf_x}{2(1-x^2)(1+f)} +\frac{f}{(1-x^2)^2}=\frac{2 q^2(1+f)}{C^2 (1-x^2)^3},\label{nu}
\end{equation}
where $f_x$ represents derivative with respect to $x$. To solve equation (\ref{nu}), we introduce a new variable
\begin{equation}
e^\nu (1+f)^{-\frac{1}{4}} = \psi(x),\label{psi}
\end{equation}
and rewrite equation (\ref{nu}) as
\begin{eqnarray}
\frac{d^2\psi}{dx^2}+\left[\frac{f_{xx}}{4(1+f)}-\frac{5f^2_x}{16(1+f)^2}+\frac{xf_x}{2(1-x^2)(1+f)}\right.\nonumber\\
\left.+\frac{f}{(1-x^2)^2}-\frac{2q^2(1+f)}{C^2(1-x^2)^3}\right]\psi=0.\label{trans}
\end{eqnarray}

Now let's try to superimpose charge distribution on the Schwarzschild solution. This suggests that $\psi(r)$ should be a linear function and when charge distribution is switched off; i.e. $f=0$. Thus, we demand $d^2\psi/dx^2 = 0$ giving us
\begin{equation}
q^2(x) = \frac{C^2(1-x^2)^3f_{xx}}{8(1+f)^2}-\frac{5C^2(1-x^2)^3f^2_x}{32(1+f)^3}+\frac{C^2x(1-x^2)^2f_x}{4(1+f)^2}+\frac{C^2(1-x^2)f}{2(1+f)},\label{Elec}
\end{equation}
and
\begin{equation}
\psi = a -b x.\label{psisol}
\end{equation}
Note that the above solution is obtained for the particular choice (\ref{Elec}) which in terms of the radial parameter $r$ takes the form
\begin{equation}
q^2(r) = \frac{r^4(C^2-r^2)f''}{8C^2(1+f)^2}-\frac{5r^4(C^2-r^2)f'^2}{32C^2(1+f)^3}-\frac{r^3(3C^2-2r^2)f'}{8C^2(1+f)^2}+\frac{r^2f}{2(1+f)}.\label{efi1}
\end{equation}
The choice of the charge distribution $q(r)$ is, therefore, motivated by the fact that it provides a simple solution which may be treated as a generalization of the Schwarzschild interior solution for an incompressible fluid as discussed below. Moreover, the choice ensures that $q(r)$ is well behaved at $r=0$ as well as at all interior points of the star.

Combining equations (\ref{psi}) and (\ref{psisol}), the unknown metric function $e^{\nu}$ is obtained as
\begin{equation}
e^\nu = (1+ f(r))^\frac{1}{4}\left(a -b\sqrt{1- \frac{r^2}{C^2}}\right),\label{lm1}
\end{equation}
where, $a$, $b$, $C$, and $k$ are constants to be determined by matching the solution to the exterior R-N metric at the boundary. Thus, the space-time metric of a  static and spherically symmetric object in the presence of an electric field is obtained as
\begin{equation}
ds_{-}^2 = -(1+ f(r))^\frac{1}{2}\left(a -b\sqrt{1- \frac{r^2}{C^2}}\right)^2 dt^2 + \frac{1+ f(r)}{1 - \frac{r^2}{C^2}} dr^2 + r^2(d\theta^2 + \sin^2\theta d\phi^2).\label{intmcomplt}
\end{equation}
Note that it reduces to the Schwarzschild solution when $f(r)=0$, and it is $f(r)$ that determines charge $q(r)^2$ in equation (\ref{efi1}). The model would be fully determined when $f(r)$ is prescribed. In the next section we shall now consider the particular choices of this function for building a model for a charged stellar object.

\subsection{Case $f(r) = 0$}
The electric field in this case vanishes and we regain the Schwarzschild interior solution for an (uncharged) incompressible fluid sphere, with $a=\frac{3}{2}\sqrt{1-\frac{R^2}{C^2}}$, $b=\frac{1}{2}$; i.e.,
\begin{equation}
ds_{-}^2 = -\left(\frac{3}{2}\sqrt{1-\frac{R^2}{C^2}}-\frac{1}{2}\sqrt{1-\frac{r^2}{C^2}}\right)^2 c^2 dt^2 + \left(1 - \frac{r^2}{C^2}\right)^{-1} dr^2 + r^2(d\theta^2 + \sin^2\theta d\phi^2),\label{intschw}
\end{equation}
where $R$ is the radius of the object and density and pressure are as given below
\begin{eqnarray}
8\pi\rho &=& \frac{3}{C^2},\label{Schden}\\
8\pi p &=& \frac{-a+3b\sqrt{1-r^2/C^2}}{C^2\left[a-b\sqrt{1-r^2/C^2}\right]}.\label{Schpres}
\end{eqnarray}
That pressure should vanish at the boundary determines the constant $R=C\sqrt{1-\frac{a^2}{9b^2}}$ which implies that we must have $a < 3b$. The central pressure takes the form
\begin{equation}
p_c = \frac{3b - a}{(a - b)C^2},
\end{equation}
which implies that we must have $a > b$. Combining the two conditions we get the bound, $ b < a < 3b$.

\subsection{Charged VT model: $f(r) =  k\frac{r^2}{C^2}$}

We must choose the function $f(r)$ such that it provides a regular, well-behaved and physically meaningful stellar model. Accordingly, we write $f(r) = k\frac{r^2}{C^2}$ which is the Vaidya-Tikekar (VT) \cite{Vaidya} ansatz for the modelling of a relativistic compact star. The $3$-hypersurface is spheroidal which becomes flat and sphere for $k = -1, 0$, respectively. The spacetime is well behaved for $r < C$ and $k > - 1$.

The energy-density, pressure and charge-density are then given by
\begin{eqnarray}
{8\pi}\rho &=& \frac{24C^4(1 + k) + 3C^2k(10 + 11k)r^2 + k^2(1 + 4k)r^4}{8(C^2 + kr^2)^3},\\
{8\pi}p &=& \frac{1}{8(C^2 + kr^2)^3[b (C^2 - r^2) - aC^2\sqrt{1 - r^2/C^2}]}\left[aC^2\sqrt{1 - r^2/C^2}[8C^4 + C^2k(22 + 9k)r^2 \right.\nonumber\\
&&\left. + k^2(9 + 4k)r^4]- b (C^2 - r^2)[24 C^4 + 9C^2k(6 + k)r^2 + k^2(25 + 4k)r^4]\right],\\
8\pi\sigma &=& \frac{\sqrt{k(C^2-r^2)}\left[3 C^4(2-k)+(4 C^2+k r^2)(7+4 k)k r^2\right]}{\sqrt{2}(C^2+k r^2)^3\sqrt{C^2(2-k)+k(7+4 k)r^2}},
\end{eqnarray}
and at the centre they take the values as
\begin{eqnarray}
8\pi\rho_c &=& \frac{3(1 + k)}{C^2},\\
8\pi p_c &=& \frac{3b - a}{(a - b)C^2}\label{cenp},\\
8\pi\sigma_c &=& \frac{3\sqrt{k(2-k)}}{\sqrt{2}C^2}.
\end{eqnarray}
Obviously, the central density remains positive for $ k > -1$ and pressure for $a > b > a/3$. For $a=b$, the central pressure diverges.

The total mass within a sphere of radius $r$ defined as
\begin{equation}
m(r) = 4\pi\int_0^r{\rho {\tilde{r}}^2 d\tilde{r}},
\end{equation}
integrates to give
\begin{equation}
m(r) = \frac{1}{16}\left[\frac{r(8k^2(1+4k)r^4-9C^4(11+7k)-kC^2(101+41k)r^2)}{8k(C^2+kr^2)^2}+\frac{9C(11+7k)\tan^{-1}{\frac{\sqrt{k}r}{C}}}{8k^{3/2}}\right]\label{mas2}.
\end{equation}
It is noteworthy that $m(r=0) = 0$.

Using Eq.~(\ref{efi1}), we have
\begin{equation}
q^2(r) =  \frac{kr^6\left[C^2(2-k) + k(7 + 4k)r^2\right]}{8(C^2 + kr^2)^3},\label{efi2}
\end{equation}
which also vanishes at the centre.

\section{Matching conditions}
\label{sec3}

The exterior spacetime of the static charged object is described by the Reissner-Nordstr\"om metric
\begin{equation}
ds_{+}^2 = -\left(1-\frac{2M}{r}+\frac{Q^2}{r^2}\right)dt^2 +\left(1-\frac{2M}{r}+\frac{Q^2}{r^2}\right)^{-1} dr^2+ r^2(d \theta^2 + \sin^2 \theta d\phi^2), \label{Vm}
\end{equation}
where $M$ and $Q$ represent the total mass and charge, respectively. The matching conditions are the continuity of $e^\nu,~ e^\mu$, and $p=0$ at the boundary $r=R$. This means $m(R)=M$ and $q(R)=Q$, and so we write
\begin{eqnarray}
m(R) = M &=& \frac{1}{16}\left[\frac{R(8k^2(1+4k)R^4-9C^4(11+7k)-kC^2(101+41k)R^2)}{8k(C^2+kR^2)^2}+\frac{9C(11+7k)\tan^{-1}{\frac{\sqrt{k}R}{C}}}{8k^{3/2}}\right],\\
q(R) = Q &=&  \sqrt{\frac{kR^6\left[C^2(2-k) + k(7 + 4k)R^2\right]}{8(C^2 + kR^2)^3}},\label{echrg}
\end{eqnarray}

The conditions explicitly take the following form at $r=R$,
\begin{eqnarray}
(1+ k n)^\frac{1}{2}(a -b\sqrt{1- n})^2 &=& 1-2u +\alpha^2 u^2,\label{bc1}\\
\frac{1 - n}{1+ k n} &=& 1-2u+\alpha^2 u^2,\label{bc2}\\
p(R) = 0 &=& a\sqrt{1-n}[8+kn(22+9k)+k^2n^2(9+4k)] \nonumber\\
&&- b(1-n)[24+9kn(6+k)+k^2n^2(25+4k)],\label{bc3}
\end{eqnarray}
where, $n=\frac{R^2}{C^2}$, $u=\frac{M}{R}$ and $\alpha^2 = \frac{Q^2}{M^2}$. Solving Eq.~(\ref{bc2}), we get
\begin{equation}
n = \frac{R^2}{C^2}  = \frac{1-y}{1+ky},\label{rsebycsq}
\end{equation}
so that
\begin{equation}
C = R\sqrt{\frac{1+ky}{1-y}},\label{const1}
\end{equation}
where $y = 1-2u +\alpha^2 u^2$. Solving Eqs.~(\ref{bc1}) and (\ref{bc3}), we obtain
\begin{eqnarray}
a &=& \frac{y[24+54kn+4k^3n^2+k^2n(9+25n)]}{16(1+kn)^2\sqrt{y\sqrt{1+kn}}},\label{const2}\\
b &=& \frac{y[8+22kn+4k^3n^2+9k^2n(1+n)]}{16\sqrt{1-n}(1+kn)^2\sqrt{y\sqrt{1+kn}}}\label{const3},
\end{eqnarray}
Thus, all the constants are expressed in terms of $k$, $M$, $R$ and $Q$.

We rewrite Eq.~(\ref{echrg}) as
\begin{equation}
\alpha^2 u^2 n =\frac{kn^3[2-k+(7+4k)kn]}{8(1+kn)^3},\label{ch1}
\end{equation}
and substitute the value of $n$ to obtain
\begin{equation}
k = \frac{4-8\alpha^2-4u\alpha^2+u^2\alpha^4-\sqrt{g(u,\alpha)}}{4-40u+8\alpha^2-4u\alpha^2+60u^2\alpha^2+u^2\alpha^4-30u^3\alpha^4+5u^4\alpha^6},\label{keq}
\end{equation}
where
$$g(u,\alpha)=16-96\alpha^2+288u\alpha^2+96u\alpha^4-456u^2\alpha^4-24u^2\alpha^6+232u^3\alpha^6-39u^4\alpha^8.$$
Note that for $\alpha =0$, we have $k=0$ while the converse is always true from Eq (39).

\subsection{Application to astrophysical objects }

A physically acceptable stellar interior solution should have the following features: (i) The density and pressure  should be positive throughout the interior of the star i.e., $\rho,~p > 0$; (ii) the pressure $p$ should vanish at some finite radial distance i.e., $p (r = R) = 0$ and (iii) the causality condition should be satisfied throughout the star which implies that $0 \leq \sqrt{\frac{dp}{d\rho}} \leq 1$.

To verify whether the above conditions are fulfilled in this model, we take the mass and radius of the pulsar $4U 1820-30$ as input parameters. The estimated mass and radius of the star are $M = 1.58~M_\odot$ and  $R = 9.1~$km, respectively \cite{Gangopadhyay}. With these values the values of the constants, for different choices of the parameter $k$,  are given in Table~\ref{tab:1}
\begin{table}[b]
\caption{\label{tab:1}
Values of the model parameters for different choices of $k$. We have assumed $M = 1.58~M_\odot$ and  $R = 9.1~$km.}
\begin{ruledtabular}
\begin{tabular}{cccc}
$k$   & $a$ & $ b$ & $C$   \\
\hline\noalign{\smallskip}
0.1 & 1.0547 & 0.5011 & 13.0983     \\
1  & 1.1299 & 0.5607 & 16.0447   \\
2 &   1.2323 & 0.6619 & 18.7246     \\ \hline
\end{tabular}
\end{ruledtabular}
\end{table}

\begin{figure}[ht]
\begin{minipage}[t]{0.495\linewidth}
\centering
\includegraphics[width=80mm]{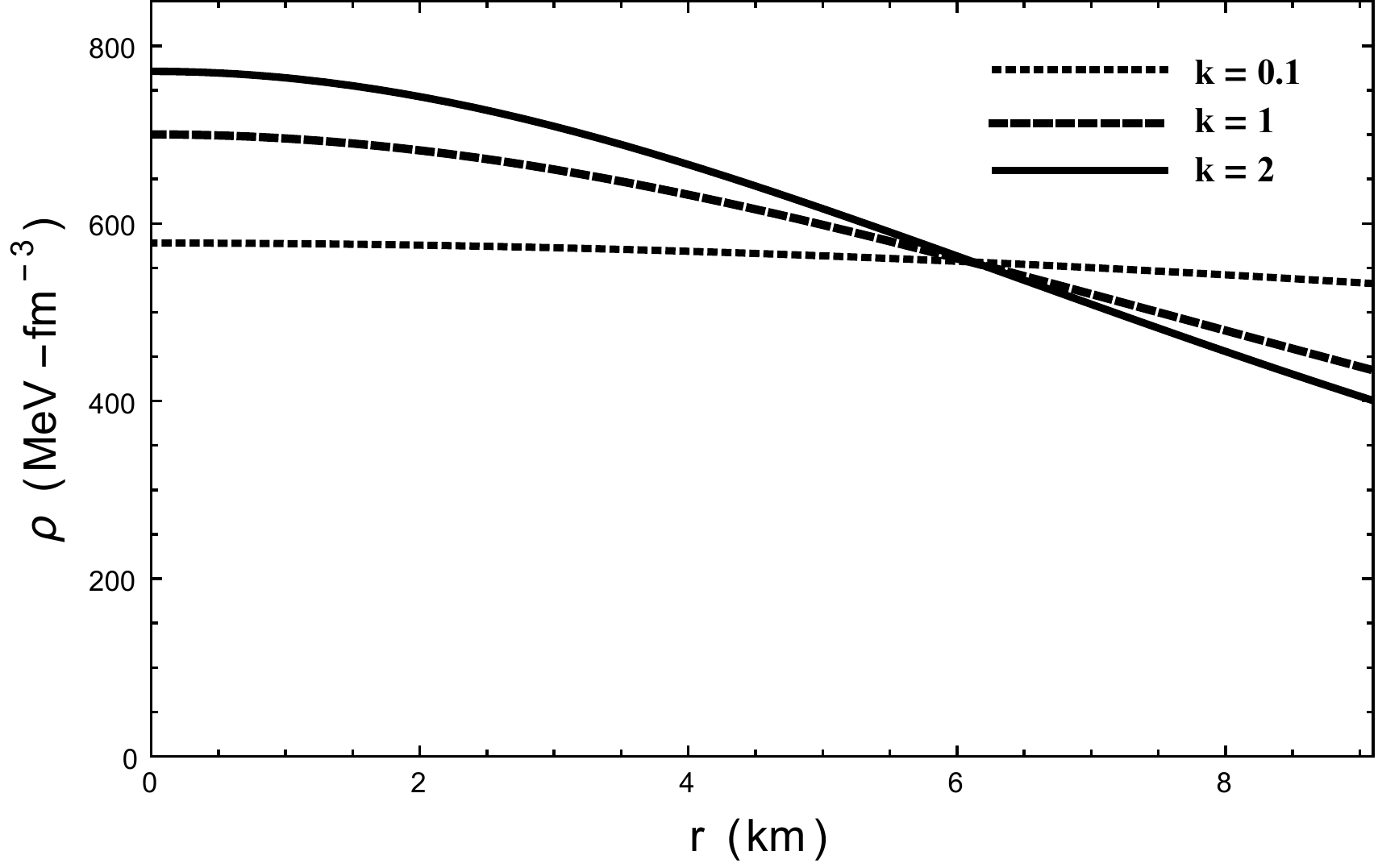}
\caption{{\small Radial dependance of density $\rho$.} }
\label{fg1}
\end{minipage}
\begin{minipage}[t]{0.495\linewidth}
\centering
\includegraphics[width=80mm]{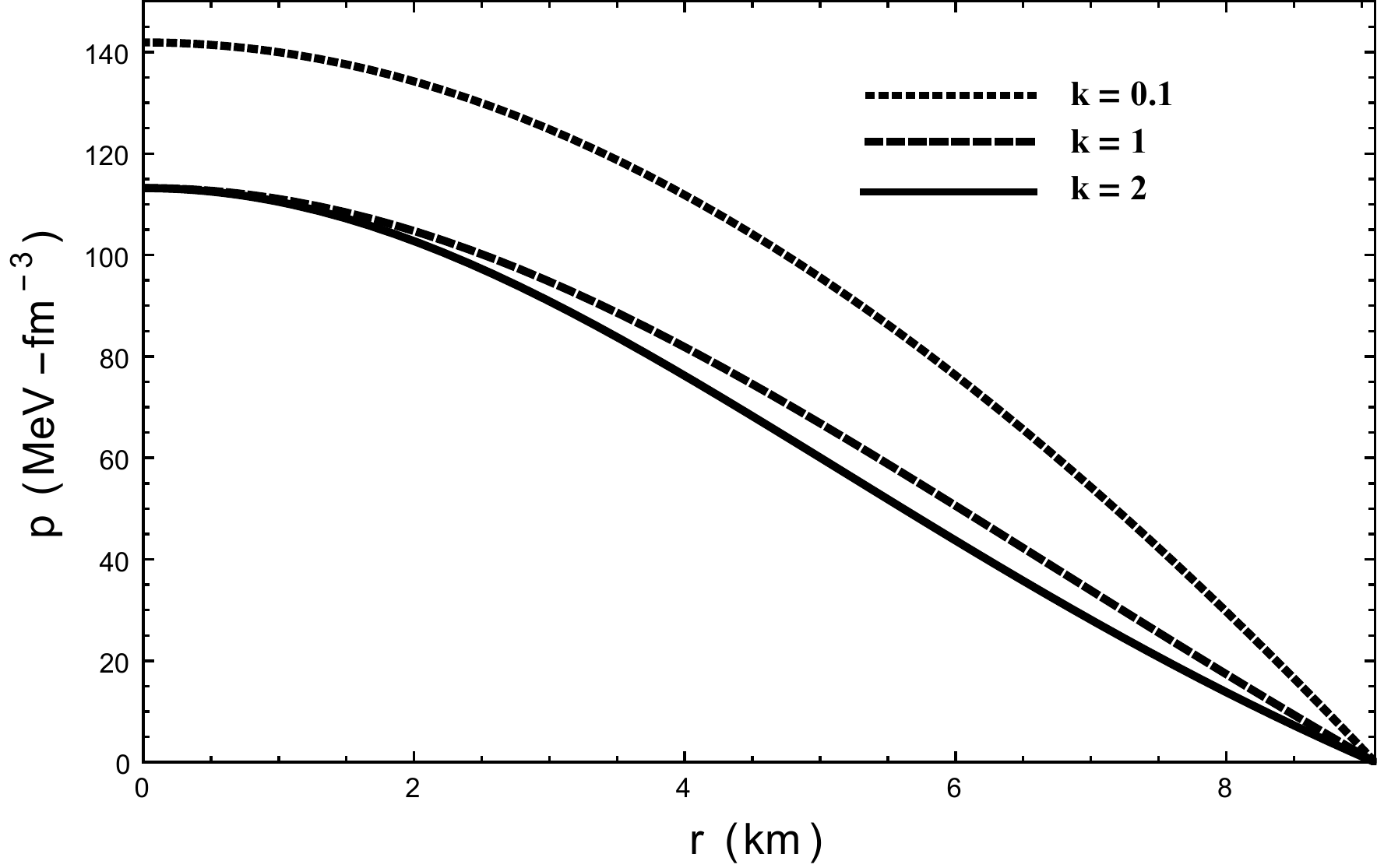}
\caption{{\small Radial dependance of pressure $p$.}}
\label{fg2}
\end{minipage}
\end{figure}

\begin{figure}[ht]
\begin{minipage}[t]{0.495\linewidth}
\centering
\includegraphics[width=80mm]{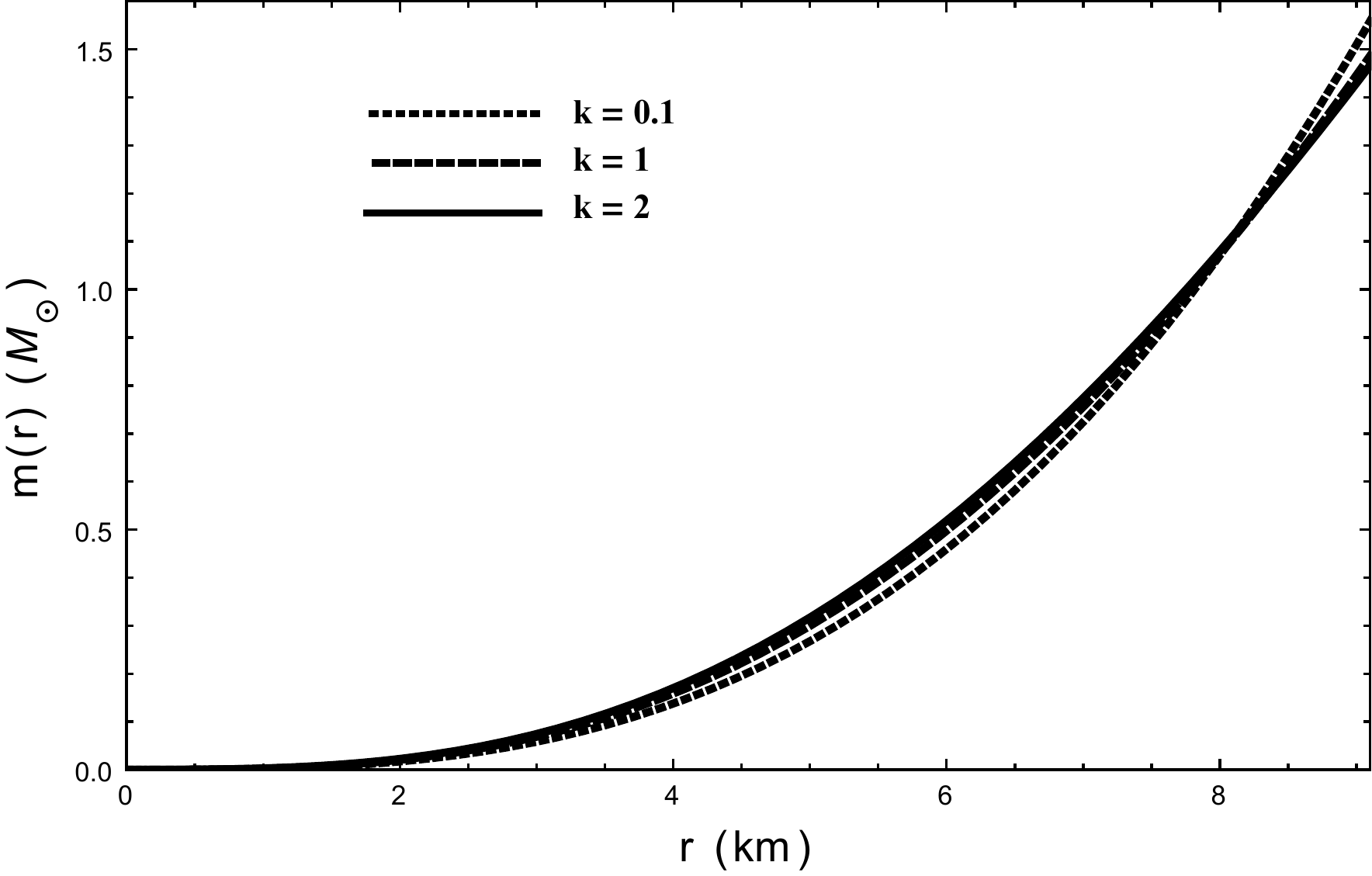}
\caption{{\small Radial variation of the mass function $m(r)$.} }
\label{fg3}
\end{minipage}
\begin{minipage}[t]{0.495\textwidth}
\centering
\includegraphics[width=80mm]{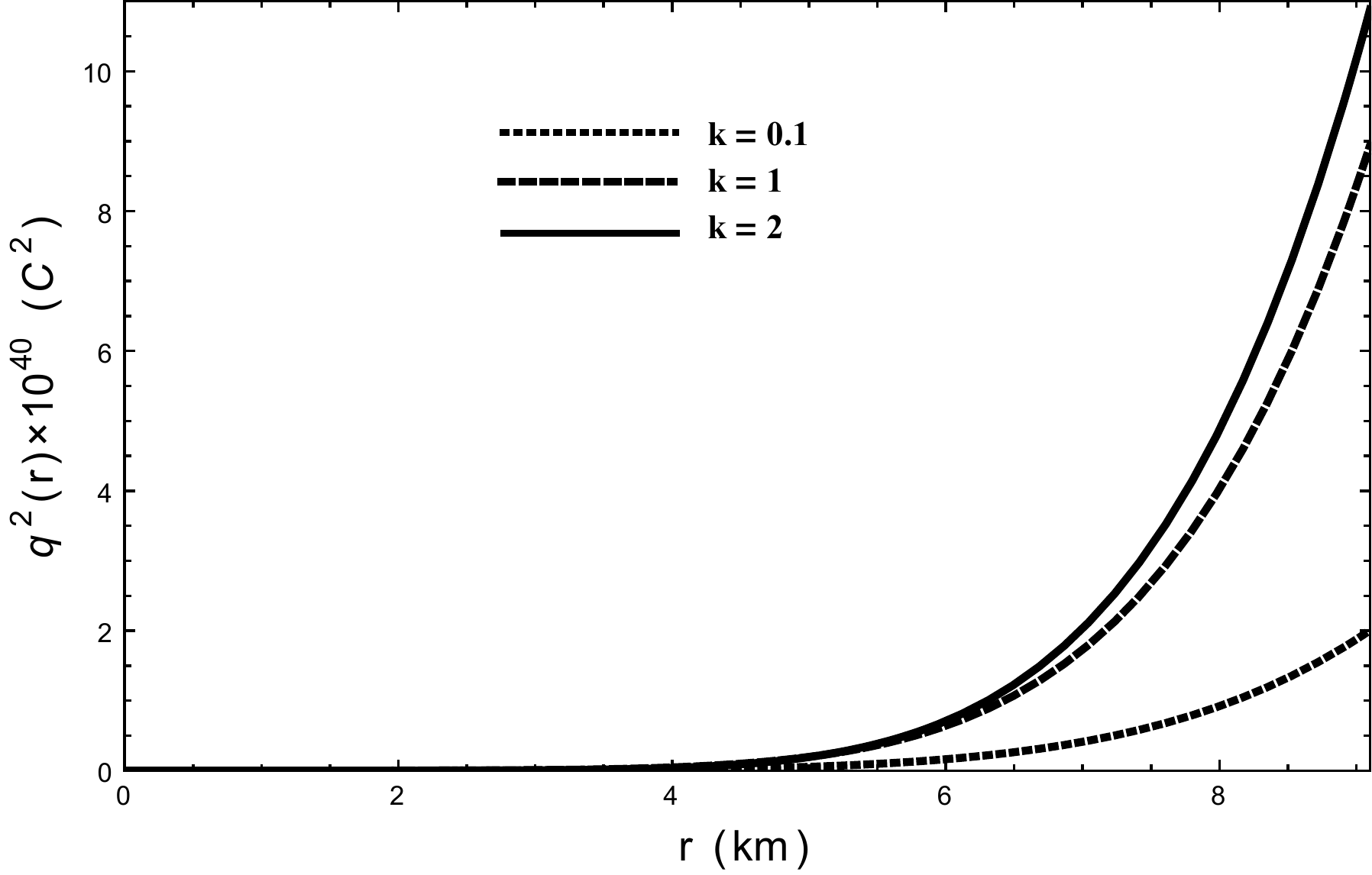}
 \caption{{\small Radial variation of charge $q^2(r)$.}}
\label{fg4}
\end{minipage}
\end{figure}

\begin{figure}[ht]
\begin{minipage}[t]{0.495\linewidth}
\centering
\includegraphics[width=80mm]{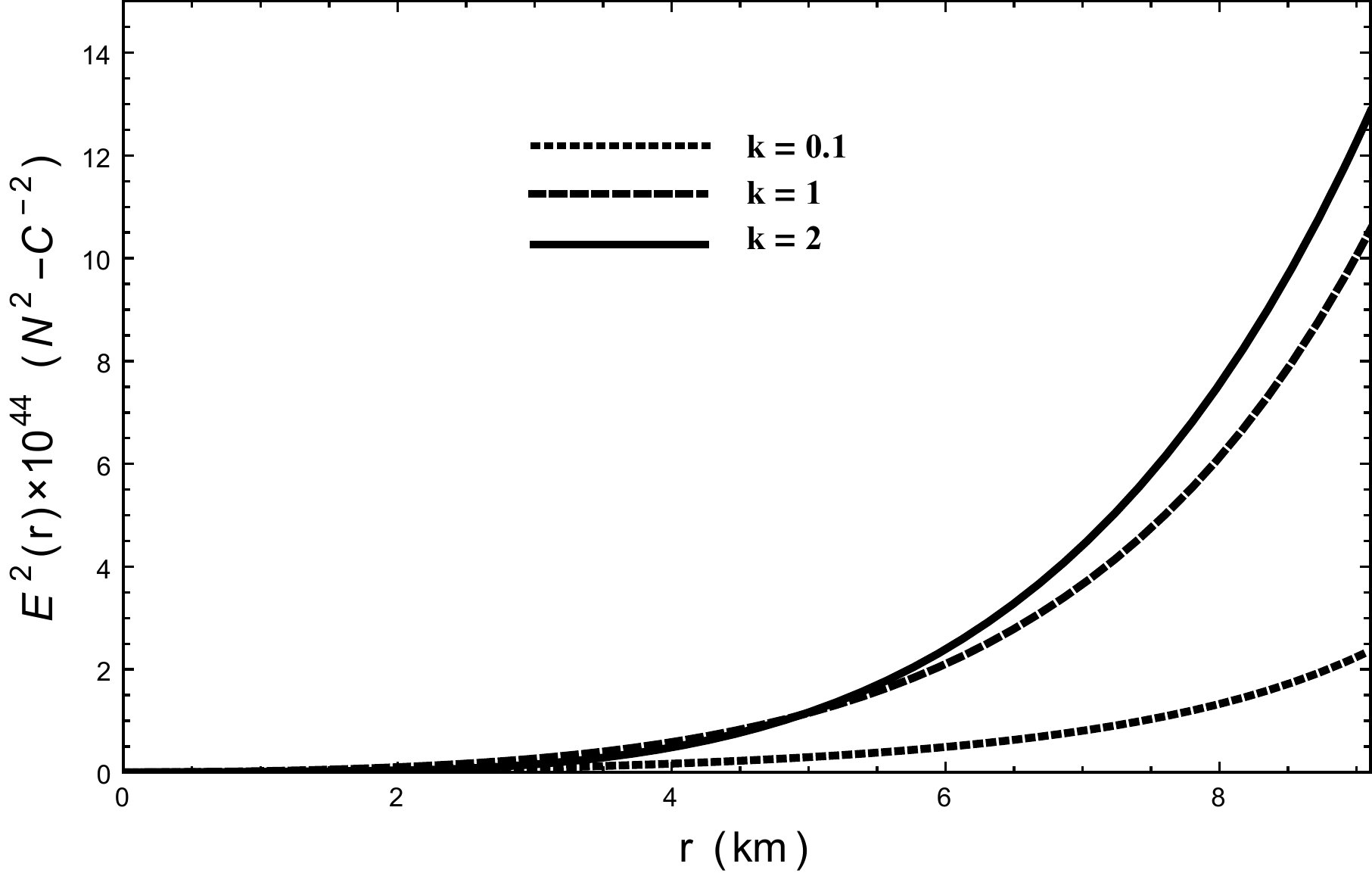}
\caption{{\small Radial variation of electric field $E^2(r)$.}}
\label{fg5}
\end{minipage}
\begin{minipage}[t]{0.495\textwidth}
\centering
\includegraphics[width=80mm]{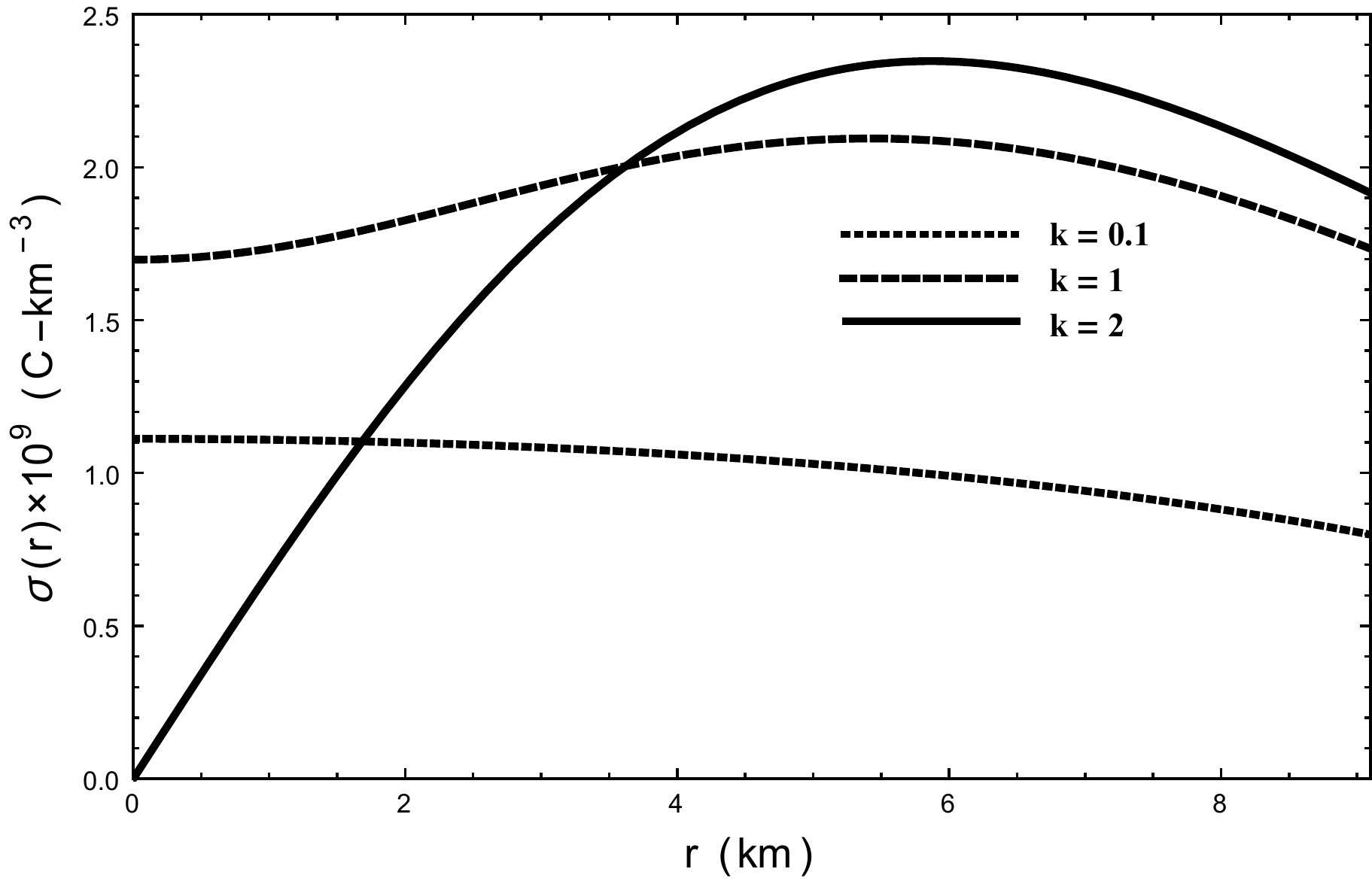}
\caption{{\small Radial variation of charge density $\sigma$.} }
\label{fg6}
\end{minipage}
\end{figure}

\begin{figure}[ht]
\begin{minipage}[t]{0.495\linewidth}
\centering
\includegraphics[width=80mm]{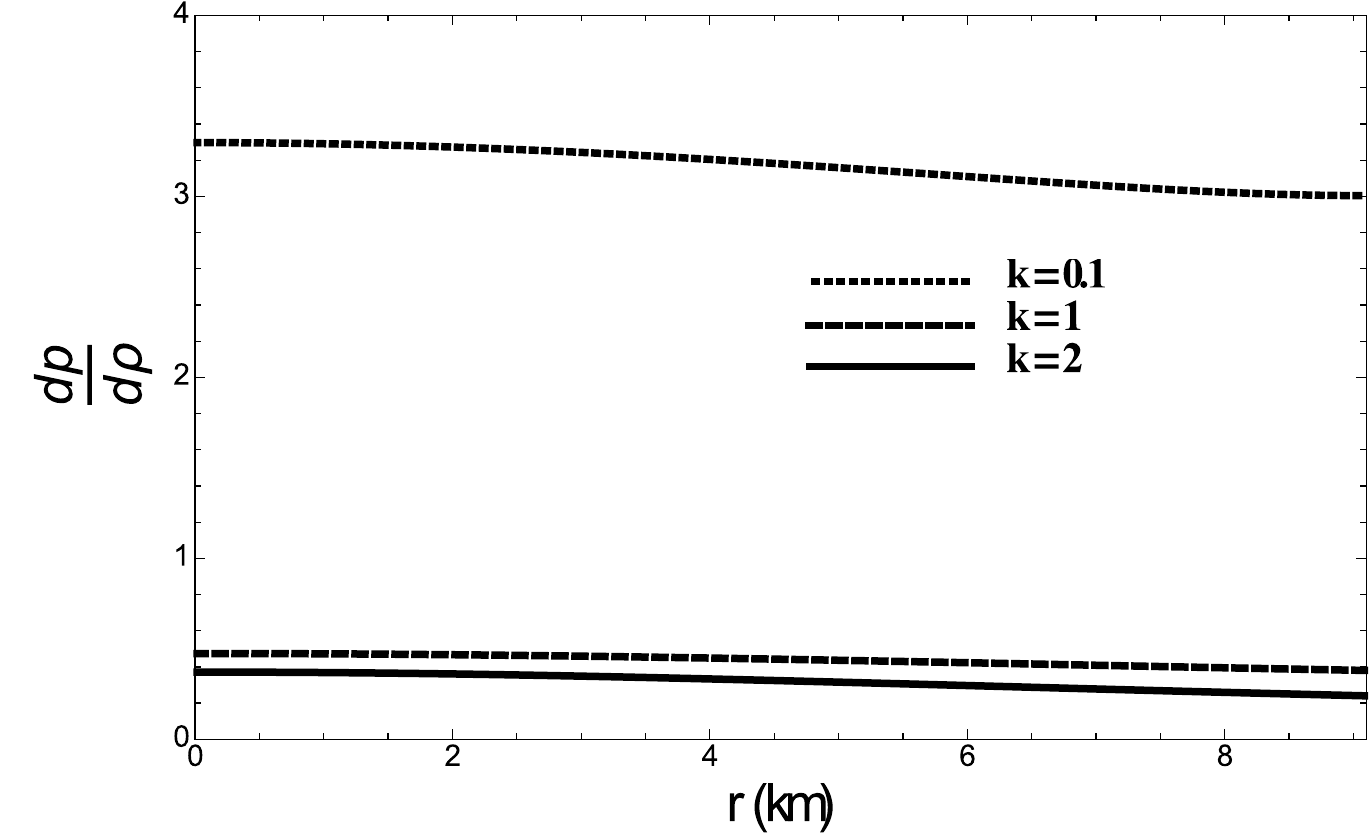}
\caption{{\small Dependance of sound speed on $k$.}}
\label{fg7}
\end{minipage}
\begin{minipage}[t]{0.495\textwidth}
\centering
\includegraphics[width=80mm]{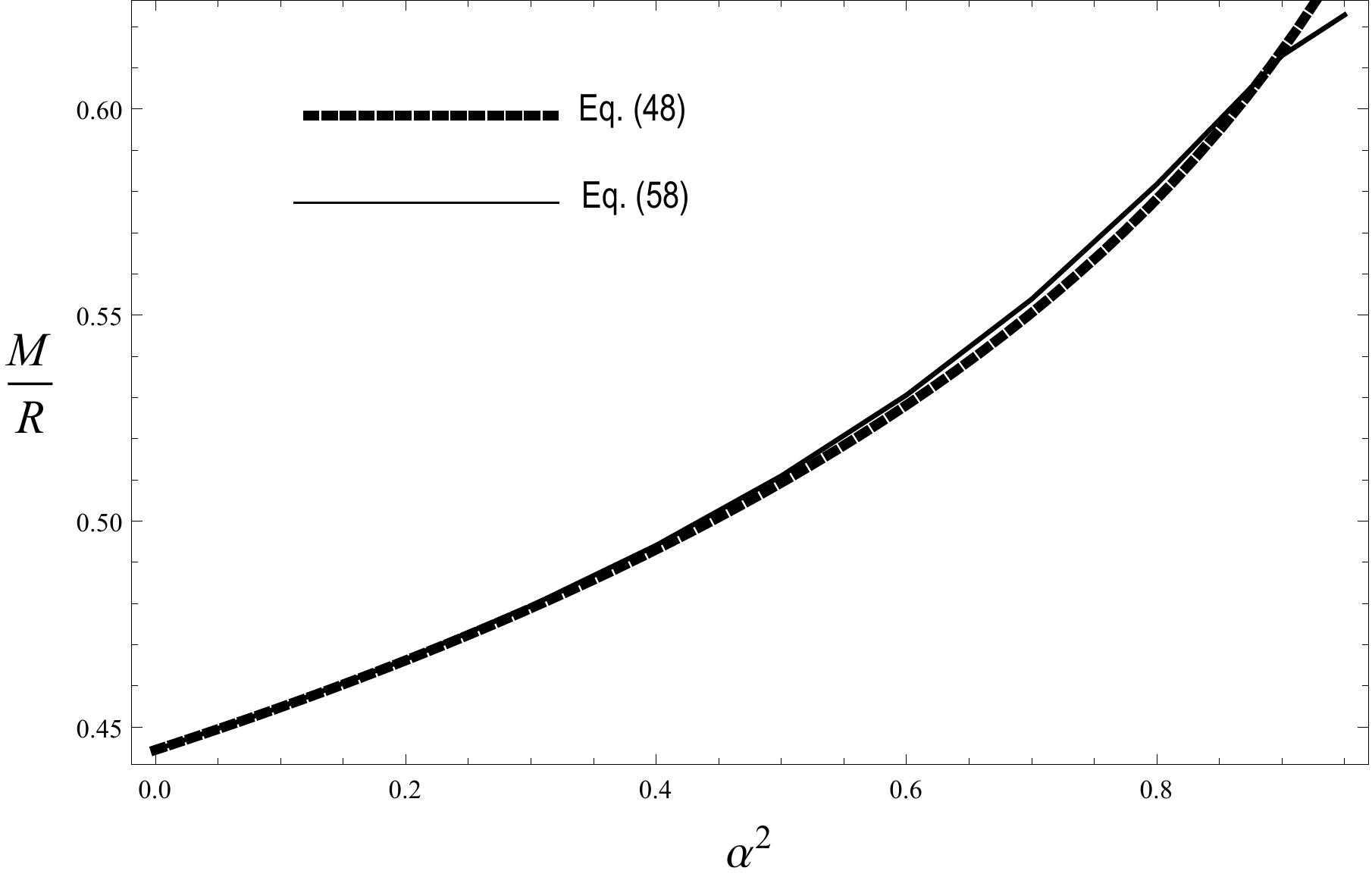}
\caption{{\small Compactness ratio $\frac{M}{R}$ plotted against $\alpha^2$ for Eqs. (48) and (58).} }
\label{fg8}
\end{minipage}
\end{figure}

It is noted that the parameter $C$, which goes inverse to energy density increases with increasing $k$; i.e. density decreases. Since $k$ is directly related to charge, which means an increase in $k$ means increase in charge. Subsequently, increase in repulsive component due to charge resulting in a decrease in fluid density. For these set of values, we show the behaviour of the physically interesting quantities in Figs.~(\ref{fg1})-(\ref{fg7}). The plots indicate that the model is regular and well-behaved at all interior points of the star.  Figs~(\ref{fg1}) and (\ref{fg2}) respectively show that density and pressure monotonically decrease with increasing radius. Note that central density is larger for larger $k$, in contrast central pressure in larger for lower $k$. That is central density and pressure are respectively largest and smallest for homogeneous fluid didtribution. The rate of fall is stronger for larger $k$, again reflecting the repulsive effect of charge. It is interesting to note that at $r\sim 6~$km, all density curves cross the uniform density straight line downwards. This clearly indicates that homogeneous distribution has the largest  density at the boundary. On the other hand, mass and charge as expected monotonically increase with the radius as shown in Figs.~(\ref{fg3})- (\ref{fg4}). Like density curves, mass curves also cross over at some $r$, and the rate of increase for larger $k$ falls down the uniform density curve. On the other hand, charge always has a stronger rate of increase with increasing $k$. Fig.~(\ref{fg5}) shows that the electric field is zero at the centre, and it monotonically increases towards the boundary and the rate of growth is stronger for larger $k$. Radial variation of the charge density is shown in Fig.~(\ref{fg6}). Note that matter density decreases with $r$ while charge density increases and attains maximum then slowly decreases. It is interesting to note that the charge density becomes maximum at a radial distance where the inhomogeneous density meets the uniform density profile.  The sound speed for different $k$ values are shown in Fig.~(\ref{fg7}) which indicates that even though for relatively higher values of $k$ the causality condition is satisfied, as we approach the constant density case ($k\sim 0.1$), sound speed becomes as expected greater than unity.

\section{Buchdahl limit}
\label{sec4}

Astrophysically it is of prime importance to find how compact a star could be, i.e. what is the upper bound on mass to radius ratio - the compactness ratio, $M/R$? From an intuitive perspective, the stiffest equation of state is of uniform density incompressible fluid which is uniquely described by the Schwarzschild interior solution. The compactness limit would be indicated by upper bound on pressure at the centre. As we have seen above, this would be the condition $a \geq b$ in Eq.~(\ref{cenp}) giving the compactness limit, $M/R \leq 4/9$. Buchdahl was first to obtain this limit under very general conditions of density and pressure being positive, the former decreasing with radius outward and at the boundary, it is matched to Schwarzschild exterior solution \cite{Buch}. The same limit is also obtained for anisotropic fluid by invoking the strong energy condition, $\rho \geq p_r + 2p_t$ \cite{Andreasson,Kara}.

For a charged object there exist more than one limits \cite{Andreasson09,Giu08,Mak01,Boehmer07} obtained for different interior distributions and equation of state. One that is closest in spirit to the uniform density case \cite{Giu08} in which it is envisaged a uniform density distribution is enveloped by a thin charged shell and the Buchdahl analogue limit reads as
\begin{equation}
\frac{M}{R} = \frac{8/9}{(1+\sqrt{1-\frac{8\alpha^2}{9}})},\, \, \alpha^2 = Q^2/M^2 \label{cbh}.
\end{equation}
It reduces to the Buchdahl limit, $M/R \leq 4/9$ when charge is switched off $Q=0$, and $M/R \leq 2/3, 8/9 < 1$ for $\alpha^2 = 1, 9/8$, respectively. Interestingly it prescribes the upper bound on charge a star could have, $\alpha^2 \leq 9/8$ which is $ > 1$. That is, a non-black hole charged object could indeed be overcharged relative to a charged black hole.

Very recently an insightful and novel prescription \cite{Dadhich20} has been proposed for the compactness limit which is given by gravitational field energy being less than or equal to half of the non-gravitational matter-energy of the object. The remarkable feature of this definition is that it is entirely determined by the unique exterior R-N metric without any reference to interior distribution, may what that be! The limit that follows is the one given above.

In the following, we would like to obtain the compactness ratio $M/R$ for our model. For a homogeneous sphere, the mass within a sphere of radius $r$ is given by
\begin{equation}
m_0(r) = \frac{4\pi\rho_0 r^3}{3},
\end{equation}
where $\rho_0 = \frac{3}{8\pi C^2}$ is the homogeneous density. Defining the compactness factor as
\begin{equation}
\phi_0(r) = \frac{m_0}{r},
\end{equation}
we write the metric potentials of the inhomogeneous sphere in the form
\begin{eqnarray}
e^{2\nu} &=& (1+ 2k\phi_0)^\frac{1}{2}\left(a -b\sqrt{1- 2\phi_0}\right)^2,\label{lm3}\\
e^{2\mu} &=& \frac{1+2k\phi_0}{1-2\phi_0}.\label{lm4}
\end{eqnarray}
We also express the physical quantities as
\begin{eqnarray}
\rho &=& \frac{\rho_0}{12(1+2k\phi_0)^3}\left[12(1+k)+3(10+11k)k\phi_0+2(1+4k)k^2\phi_0^2\right],\label{dden} \\
p &=& \frac{\rho_0\left[a(4+(22+9k)k\phi_0+2(9+4k)k^2\phi_0^2)-b\sqrt{1-2\phi_0}(12+9(6+k)k\phi_0+2(25+4k)k^2\phi_0^2)\right]}{12(1+2\phi_0)^3(-a+b\sqrt{1-2\phi_0})},\label{dpres}\\
\sigma &=& \frac{\rho_0}{3}\frac{\sqrt{k(1-2\phi_0)}\left[3(2-k)+4k\phi_0(2+k\phi_0)(7+4k)\right]}{(1+2k\phi_0)^3\sqrt{2(2-k)+4k\phi_0(7+4k)}},\label{dsig}\\
q^2(r) &=&\frac{km_0^2}{(1+2k\phi_0)^3}\left[1-\frac{k}{2}+(7+4k)k\phi_0\right]\label{dchar}.
\end{eqnarray}
Equation (\ref{dchar}) shows that charge in our model is proportional to the constant density mass $m_0$. The dimensionless parameter $k$, representing deviation from sphericity, also gets tagged into the expression of charge.

Since the departure from sphericity is expected to be small, expanding $\tan^{-1}\frac{\sqrt{k}r}{C} = \frac{\sqrt{k}r}{C} -\frac{1}{3}{(\frac{\sqrt{k}r}{C})}^3$ in equation (\ref{mas2}), we obtain the mass function as
\begin{equation}
m(r) = \frac{m_0\left[32(1+k)+(41+53k)k\phi_0-6(11+7k)k^2\phi_0^2\right]}{32(1+2k\phi_0)^2},
\end{equation}
which shows that for $k=0$ we regain $m(r) = m_0$.

At the centre $r=0$, we have
\begin{eqnarray*}
e^{2\nu} = (a-b)^2,~~ e^{2\mu} = 1,~~q^2 = 0,\\
\sigma_c = \rho_0\sqrt{k(1-\frac{k}{2})},\\
\rho_c = \rho_0(1+k),~~p_c = \frac{\rho_0(3b - a)}{3(a-b)}.
\end{eqnarray*}
The regularity of $\sigma_c$ demands that we must have $k \leq 2$. We also note that $p_c \rightarrow \infty$ for $a=b$. In other words, for a stellar configuration, we must have $a > b$ and consequently, the upper bound on the compactness vis-a-vis Buchdahl type limit can be obtained by setting $a=b$.

Now, imposing the condition $a \geq b$ in Eqs.~(\ref{const2}) and (\ref{const3}), we obtain
\begin{eqnarray}
\sqrt{\frac{(1+k)y}{1+ky}}\left[-24+6k(-4-2u+\alpha^2 u^2)+k^2u(-18u(20+9\alpha^2)-20\alpha^2 u^2+5 \alpha^4 u^3)\right]\nonumber\\
+8-2k(-4-6u+3\alpha^2 u^2)-k^2u\left(-18u(20+9\alpha^2)-20\alpha^2 u^2+5\alpha^4 u^3\right) \geq 0\label{maseq}
\end{eqnarray}

In the absence of any simple solution of Eq.~(\ref{maseq}), let us first consider the case $k=0$. Substituting $k=0$, in Eq.~(\ref{echrg}), we obtain $Q=0$ which implies that we must set $\alpha=0$ in this case. Note that in the uncharged case ($k=0$), the exterior Reissner-Nordstr\"om metric should be replaced by the Schwarzschild solution by setting $Q=0$. Therefore, substituting $k = 0 = \alpha$ in Eq.~(\ref{maseq}), we obtain
\begin{equation}
3\sqrt{1-2u} \geq 1,
\end{equation}
which readily provides the Buchdahl limit $u (=M/R) \leq 4/9$.

For $k \neq 0$, substituting the value of $k$ from Eq.(\ref{keq}) into Eq. (\ref{maseq}), and solving it numerically, one can find the upper bound on $u = \frac{M}{R}$ for a given charge to mass ratio, $\alpha^2=\frac{Q^2}{M^2}$. The results are shown in Table~{\ref{tab:2} and Fig.~(\ref{fg8}). Alongside Eq. (58) we also plot Eq. (48) clearly indicating how beautifully our model is coasting along with the exact Buchdahl bound for charged object.

\begin{table}[b]
\caption{\label{tab:2}
Compactness $\frac{M}{R}$ for different values of $\alpha^2$.}
\begin{ruledtabular}
\begin{tabular}{|c|c|c|c|c|c|c|c|}
$\alpha^2$ & 0 & 0.1 & 0.3 & 0.5 & 0.7 & 0.9 & 0.95    \\ \hline
$\frac{M}{R}$ & $\frac{4}{9}$ & 0.4548 & 0.4794  & 0.5109 &  0.5538  & 0.6130  & 0.6228 \\  \hline
\end{tabular}
\end{ruledtabular}
\end{table}

\subsection{Approximation method}

To reduce the complexity of the equations, we set $k = k\epsilon$, where $0 < \epsilon << 1$. This is a reasonable assumption as the departure from sphericity denoted by $k$ is expected to be small. In this case, retaining terms up to $\mathcal{O}(\epsilon)$, we use (\ref{ch1}) to obtain
\begin{equation}
\alpha^2 = \frac{4-4k+12ku-\sqrt{-80k^2u^2+(-4+4k-12ku)^2}}{10ku^2}.\label{alp}
\end{equation}
Eq.~(\ref{alp}) ensures that we have $\alpha=0$ for $k=0$. Inserting the value of $\alpha^2$ in Eq.~(\ref{maseq}) and neglecting terms $\mathcal{O}(\epsilon^2)$, we obtain
\begin{equation}
3\sqrt{1-2u}+\frac{3ku^2}{2\sqrt{1-2u}}=1.\label{appx1}
\end{equation}
Even though this is an approximate equation, a neat solution of the above is not available. Nevertheless, numerical solution of the above equation provides a generalization of the Buchdahl limit in the case of a charged sphere. It is not difficult to show that for $k=0$, the equation yields the Buchdahl limit $u \leq 4/9$.

To find an analytic solution of (\ref{maseq}) in the extreme case $a=b$ (when the central pressure diverges), we make use of equation (\ref{keq}) and obtain a truncated equation up to the order $\mathcal{O}(\epsilon)$ as
\begin{equation}
3\sqrt{1-2u+u^2\alpha^2}=1,\label{maseq2}
\end{equation}
which readily yields
\begin{equation}
u = \frac{3\alpha^2+\sqrt{9-8\alpha^2}}{3\alpha^2}.\label{final}
\end{equation}
Rationalizing the above, we finally obtain a charged analogue of the Buchdahl limit given by
\begin{equation}
u= \frac{M}{R} = \frac{8/9}{(1+\sqrt{1-\frac{8\alpha^2}{9}})}.
\end{equation}
The above result provides an upper bound on $\alpha^2 \leq 9/8$ and $u \leq 8/9 < 1$. Non black hole object would always have radius larger than the black hole. For $\alpha^2 = 0$, we regain $u \leq 4/9$. It is remarkable to note that by making use of a different technique and a specific model, we have been able to obtain the desired upper bound on the compactness of a charged object - a charged Buchdahl limit.

\section{Discussion}
\label{sec5}

By employing the Vaidya-Tikekar metric ansatz we have constructed a generalization of the homogeneous density distribution in which the parameter $k$ gets coupled to charge distribution. When charge is set to zero, the solution goes over to the Schwarzschild uniform density fluid sphere. That means it is kind of charging the Schwarzschild uniform density solution.

This has facilitated computation of the charged analogue of the Buchdahl compactness limit in Eq.~(\ref{maseq}). In Fig.~\ref{fg8}, the compactness ratio $M/R$ is plotted for Eqs.~(\ref{cbh}) and (\ref{maseq}), the former is the Buchdahl bound as found in Refs.~\cite{Giu08} and \cite{Dadhich20} while the latter is computed for the present model. It is remarkable to see that how the one due to Eq.~(ref{maseq}) coasts beautifully that due to Eq.~(\ref{cbh}). 

Another worth noting feature is that where charge density (Fig.~\ref{fg6}) attains the maximum value where energy density crosses the uniform density line (Fig.~\ref{fg1}). This indicates that charge density increases while energy density decreases with radius until the latter crosses the uniform density line. Then the former attains maximum value and begins decreasing. This means for radius greater than the one where energy density becomes less than the uniform density value, charge density also begins decreasing. It is interesting to see how the behaviour of charge density is linked to the fact whether energy density is greater or less than the uniform density value.

It is remarkable that a charged object could have $\alpha^2 = Q^2/M^2 \leq 9/8 > 1$; i.e. it could be overcharged relative to a charged black hole. The question then arises, is it possible to construct models with $1\leq \alpha^2 \leq 9/8$ - an explicit example of overcharged object? It would be interesting to construct such a model and that's what we would like to take up next in a separate investigation.

\begin{acknowledgments}
RS gratefully acknowledges support from the Inter-University Centre for Astronomy and Astrophysics (IUCAA), Pune, India,  under its Visiting Research Associateship Programme.
\end{acknowledgments}

\end{document}